\documentclass[useAMS,usenatbib]{mn2e}
\usepackage{graphicx}
\usepackage{natbib} 
\usepackage{abbrev}

\title[Galaxy Alignments in Clusters at $z>0.5$]{Galaxy Alignments in  Very X-ray Luminous Clusters at $z>0.5$}
\author[Chao-Ling Hung \& Harald Ebeling]
{Chao-Ling Hung$^1$\thanks{Email: clhung@ifa.hawaii.edu} \&
Harald Ebeling$^1$
\\
$^{1}$Institute for Astronomy, University of Hawaii, 2680 Woodlawn Drive, Honolulu, HI 96822, USA}
\begin{document}

\date{Accepted 2012 January 12. Received 2012 January 11; in original form 2011 December 10}

\pagerange{\pageref{firstpage}--\pageref{lastpage}} \pubyear{2002}

\maketitle

\label{firstpage}

\begin{abstract}
We present the results of a search for galaxy alignments in $12$ galaxy clusters at $z>0.5$, a statistically complete subset of the very X-ray luminous clusters from the MAssive Cluster Survey (MACS). Using high-quality images taken with the Hubble Space Telescope (HST) that render measurement errors negligible, we find no  radial galaxy alignments within 500 kpc of the cluster centres for a sample of 545 spectroscopically confirmed cluster members. A mild, but statistically  insignificant trend favouring radial alignments  is observed within a radius of 200 kpc and traced to galaxies on the cluster red sequence. Our results for massive clusters at $z>0.5$ are in stark contrast to the findings of previous studies which find highly significant radial alignments of galaxies in nearby clusters at $z\sim0.1$ out to at least half the virial radius using imaging data from the SDSS. The discrepancy becomes even more startling if radial alignment becomes more prevalent at decreasing clustercentric distance, as suggested by both our and previous work. We investigate and discuss potential causes for the disparity between our findings based on HST images of clusters at $z>0.5$ and those obtained using groundbased images of systems at $z\sim 0.1$. We conclude that the most likely explanation is either dramatic evolution with redshift (in the sense that radial alignments are less pronounced in dynamically younger systems) or the presence of systematic biases in the analysis of SDSS imaging data that cause at least partly spurious alignment signals. 
 \end{abstract}

\begin{keywords}
galaxies: clusters: general; galaxies: evolution; galaxies: kinematics and dynamics
\end{keywords}

\section{Introduction}\label{sec:intro}
Galaxy and cluster orientations provide crucial insights into the processes governing cluster formation and evolution.
On large scales, the distributions of galaxies in clusters are found to align with the direction toward their neighbours  within  distances of tens of Mpc \citep{Binggeli1982}, a tendency that is explained by a preference for clustering and merging along  large-scale filamentary structures  \citep{West1995}. On smaller scales, intrinsic galaxy alignments generated during galaxy formation are observed on scales of a few Mpc \citep{Lee2007} with different origins, however, for  elliptical and spiral galaxies. Similar to the physical mechanisms behind cluster alignments, anisotropic accretion of material onto galaxy halos along filamentary structures can result in the alignment of elliptical galaxies \citep{West1994}. By contrast, the alignment of spiral galaxies is believed to be caused by correlations between their spin axes originating from initial large-scale tidal fields \citep{Pen2000}.

In dense cluster environments, extensive evidence is found of alignments between the brightest cluster galaxies (BCGs) and their host-cluster halos \citep{Binggeli1982}, an effect that is again caused by dynamical connections to large-scale filamentary structures. For cluster members other than the BCGs, any intrinsic alignment might be expected to be destroyed by their gravitational interaction with the host-cluster halos on a time scale of a few orbits \citep{Coutts1996}. However, using Sloan Digital Sky Survey (SDSS) data in a study of a sample of 85 X-ray selected clusters at $0.02<z<0.23$, \citet{Pereira2005} (hereafter PK05) find a tendency for the major axes of cluster galaxies to be aligned with the radius vector pointing toward the cluster centre. They propose that this radial alignment  is the result of tidal torquing exerted by the host cluster halo \citep{Pereira2008,Pereira2010}. \citet{Faltenbacher2007} (hereafter F07) confirm the radial alignment signal for a sample of galaxy groups at $0.02<z<0.2$ selected from the SDSS group catalogue, and  further find a trend for  the signal to be more significant near the cluster centre and for redder galaxies.

So far almost all searches for galaxy alignments have been conducted for relatively nearby clusters owing to the limited depth of readily available ground-based imaging data. If a study akin to PK05 could be conducted at significantly higher redshift, the presence or absence of galaxy alignments in distant clusters could shed light on the temporal evolution of the effect and thus on the dynamical timescales of the physical mechanisms at work.
Unfortunately, measuring galaxy orientations from SDSS data is extremely challenging already at low to moderate redshifts due to the poor seeing and insufficient pixel sampling.
High-quality images are thus essential in obtaining precise galaxy orientation measurements, specifically at high redshift.

In this paper, we discuss our study of galaxy alignments in 12 X-ray luminous clusters at $z >0.5$ from the Massive Cluster Survey \citep[MACS;][]{Ebeling2007} as a first step toward investigating the evolution of galaxy alignments with lookback time. All  12 clusters were observed with the Advanced Camera for Surveys (ACS) aboard the Hubble Space Telescope (HST). We describe our target clusters, the HST observations, and image reduction procedures in \S 2.  Cluster membership determination and galaxy shape measurements are described in \S 3. Results on galaxy alignments are presented in \S 4.1 and \S 4.2, and possible systematics are examined in \S 4.3. We interpret our results and compare them with those of studies in nearby clusters in \S 5; a conclusion is given in \S 6.

AB magnitudes are used throughout. We adopt the concordance $\Lambda$CDM cosmology with $H_0=70$ km s$^{-1}$ Mpc$^{-1}$, $\Omega_{M}=0.3$ and $\Omega_{\Lambda}=0.7$.

\section{Twelve MACS Clusters at $z > 0.5$}

MACS provides a statistically complete sample of the most X-ray luminous clusters at $z>0.3$ \citep{Ebeling2001}.
\citet{Ebeling2007} present a subsample of the 12 most distant MACS clusters at $z>0.5$, with 11 of them at $0.5<z<0.6$ and MACS J0744.8+3927 at $z=0.698$. Like all MACS clusters, these systems are very massive \citep[$M\sim 10^{14}-10^{15} M_{\odot}$;][]{Mantz2010}. Being purely X-ray selected, this sample is highly diverse with respect to optical appearance and dynamical state. For example, MACS\,J0025.4--1222 consists of two substructures of nearly equal mass with the hot gas residing in between them, showing it to be a post-collision merger\citep{Bradac2008,Mann2011}. Similarly, in-depth analyses of MACS\,J0717.5+3745 and MACS\,J1149.5+222 show highly complex structures and dynamics \citep{Ma2009, Smith2009}. On the other hand, MACS\,J1423.8+2404 features a prominent cool core and a single dominant cD galaxy, implying that it is close to fully virialized \citep{Limousin2010}.

\subsection{HST Imaging}

\begin{figure*}
 \centering
  \includegraphics[width=0.9\textwidth]{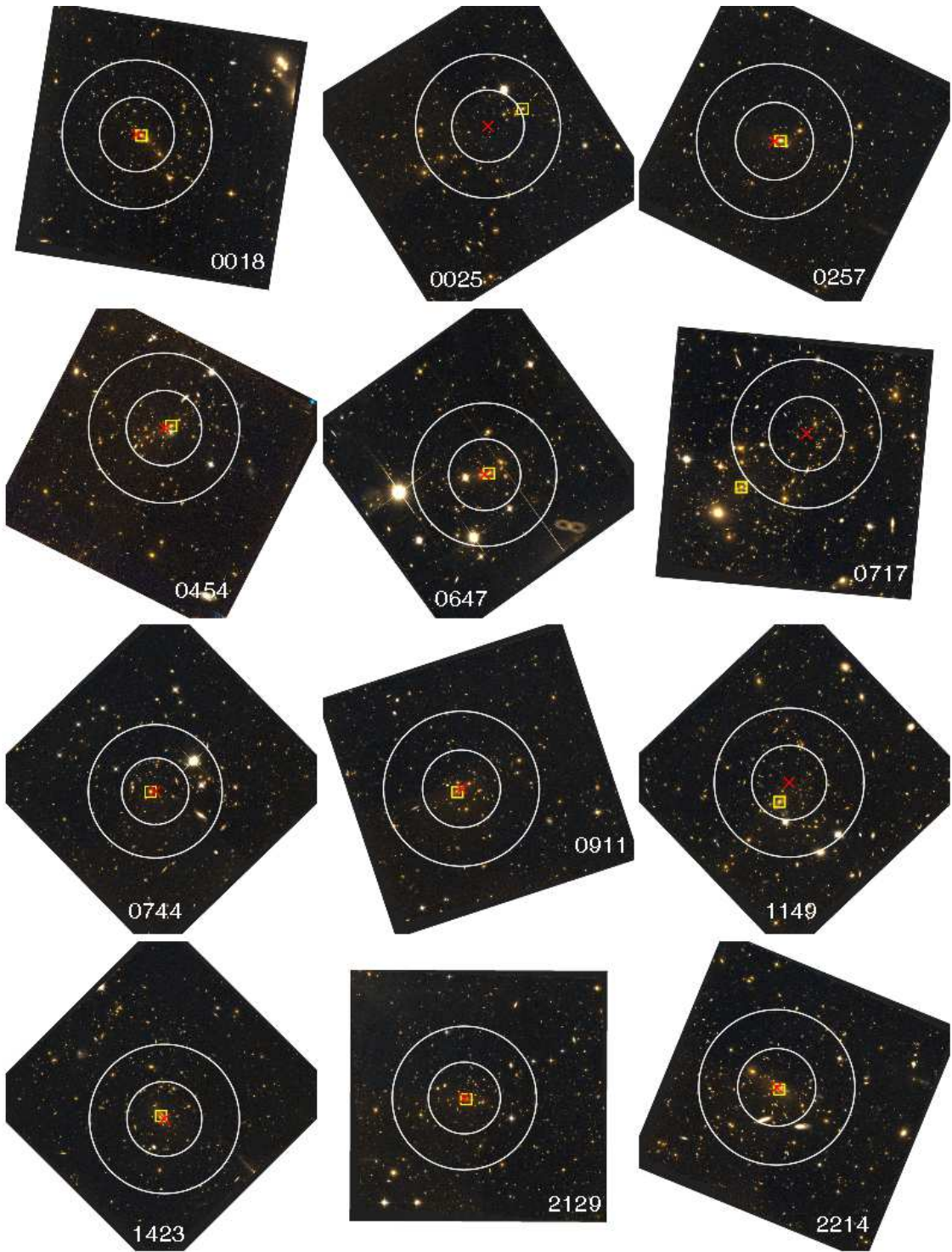}
\caption{Three-color images of our 12 MACS clusters with F555W, (F555W+F814W)/2 and F814W used for the blue, green and red channel, respectively. Clusters are shown in order of ascending right ascension from left to right and top to bottom. In each cluster field, the red cross indicates the location of the peak of the X-ray surface brightness, while the yellow box marks the chosen BCG. The inner and outer white circles have a radius of $200$ and $400$ kpc at the cluster redshift, respectively.}
\label{fig:hstimg}
\end{figure*}

All 12 target clusters were imaged with the ACS (Wide Field Channel) on HST (PID: 9722, 10703, 11560, PI: H. Ebeling; PID: 10493, PI: A. Gal-Yam). The field of view of ACS/WFC ($3.5\times 3.5$ arcmin$^2$) corresponds to approximately $1.23\times 1.23$ Mpc$^2$ at $z{\sim}0.55$ and is thus well matched to the physical size of the cluster cores. Each cluster field was imaged in the F555W and F814W filters for about 4500 seconds. Bad-pixel masking, geometric distortion correction, cosmic ray rejection, image stacking and resampling were performed on the flat-fielded images using the MultiDrizzle program \citep{Koekemoer2002}. For MACS\,J0018.5+1626, observed in November 2010, additional Charge Transfer Efficiency (CTE) corrections were applied to the data using the pixel-based CTE correction code \citep{Anderson2010} before further reduction with the MultiDrizzle program. We applied the optimized resampling scales of $0.03^{\prime \prime}$, a Gaussian drizzle kernel, and pixfrac $=0.8$ to avoid aliasing in the Point Spread Function \citep{Rhodes2007}. Pseudo-color HST/ACS images of all 12 clusters are shown in Figure~\ref{fig:hstimg}. Object detection and photometry were performed on the reduced ACS images using SExtractor \citep{Bertin1996}.
We removed the objects with SExtractor FLAGS larger than 4 as well as point-like and spurious sources identified from their location in the distribution of peak surface brightness versus magnitude. 

\section{Analysis}

\subsection{Cluster Membership Determination}\label{sec:membership}

The most accurate way to determine cluster membership is via spectroscopic observations of galaxies in our target fields. Spectroscopically determined redshifts are available for $\sim$1150 objects from C.-J. Ma et al. (2011, in preparation), \citet{Ma2008} and the NASA/IPAC Extragalactic Database. Galaxies with a radial velocity within $3 \sigma$ (where $\sigma$ indicates the dispersion in the velocity distribution) of the systemic cluster velocity are taken to be cluster members (hereafter \textit{sample A}), while those with a radial velocity above or below the $3 \sigma$ threshold are selected as background and foreground galaxies, respectively.

\begin{table}
 \caption{Numbers of cluster members in the 12 MACS clusters within the ACS field of view for various subsamples:
 spectroscopically confirmed (A), red-sequence selected (B),
 both spectroscopically and colour-selected (A+B) and spectroscopically but not colour selected (A-B).}
 \label{tab:count}
 \begin{tabular}{@{}lcccc}
 \hline
 \hline
MACS name & n(A) & n(B) & n(A+B) & n(A-B) \\
  \hline
 MACSJ0018.5$+$1626  &100 &233 & 62 & 38\\
MACSJ0025.4$-$1222   & 74 & 223 & 62 & 12 \\
MACSJ0257.1$-$2325  &  50& 142 & 33 & 17 \\
MACSJ0454.1$-$0300   & 75 & 159 & 52 & 23 \\
MACSJ0647.7$+$7015   & 30& 71 & 12 & 18 \\
MACSJ0717.5$+$3745   & 120& 316 & 97 & 23 \\
MACSJ0744.8$+$3927  & 65 & 160 & 46 & 19 \\
MACSJ0911.2$+$1746  & 63& 162& 46& 17 \\
MACSJ1149.5$+$2223  & 80 & 267& 58& 22\\
MACSJ1423.8$+$2404   & 47&  127&  38& 9 \\
MACSJ2129.4$-$0741   & 82& 173 & 54 & 28 \\
MACSJ2214.9$-$1359   & 86& 209& 67& 19 \\
\hline
Total &       872 & 2242 & 627  &245\\
\hline
        
 \end{tabular}

\end{table}

\begin{figure}
  \includegraphics[width=0.5\textwidth]{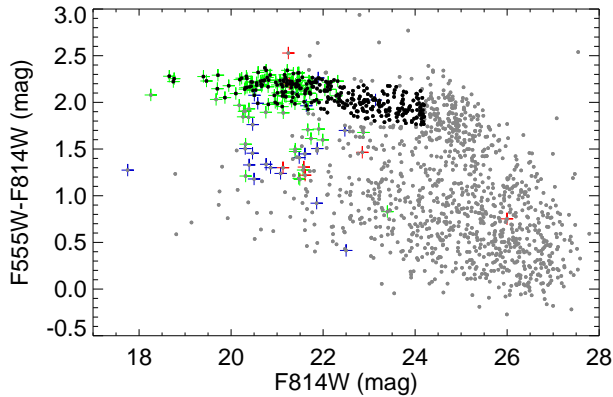} 
\caption{Colour magnitude diagram of MACS\,J0717.3+3745. The grey dots represent all galaxies in the field, and the black dots are red-sequence selected cluster members. The crosses mark galaxies with redshift information where green, blue, and red indicate spectroscopically confirmed members, foreground, and background galaxies, respectively.}
\label{fig:cmd}
\end{figure}

To increase the sample size, we alternatively select cluster members based on galaxy colour. For each cluster we identify the red sequence from a colour-magnitude diagram (F555W--F814W versus F814W;  see Figure~\ref{fig:cmd} for an example). The red sequence is formed by old elliptical galaxies whose spectra show similar $D_{4000}$ breaks resulting from photospheric absorptions of heavy elements \citep{Bower1992}. Galaxies falling within $\pm 2 \sigma$ of the red sequence (here $\sigma$ is the dispersion in the red sequence along the F555W--F814W axis) and featuring magnitudes of $F814W<24$ mag are classified as cluster members (hereafter \textit{sample B}).
The numbers of cluster members in each sample and for each cluster are listed in Table~\ref{tab:count}.

\subsection{Galaxy Orientation Measurement}

\begin{figure}
 \centering
 \includegraphics[width=0.3\textwidth]{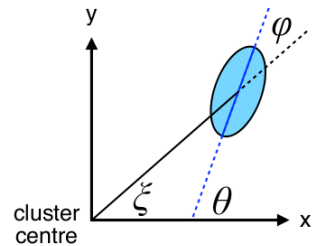} 
\caption{Definition of  $\theta$, $\xi$, and  $\phi$.}
\label{fig:phi}
\end{figure}

Galaxy orientations are measured with SExtractor from the F814W images. We use the windowed positional parameters from SExtractor with all derived quantities based on the weighted second-order brightness moments,

\begin{equation}
 Q_{ij}=\int d^{2}\theta W_{r_g}(|\theta|) \theta_{i} \theta_{j} I(\theta),
 \end{equation}

where $W_{r_g}(|\theta|)$ indicates a 2-dimensional Gaussian with a dispersion of $r_g$, and $I(\theta)$ represents the brightness distribution. The windowed positional parameters are derived adopting $r_g={\rm FWHM}/\sqrt{8\ln2}$.
The impact of the precise choice of $r_g$ is discussed in \S 4.3.2. We apply a Gaussian weighting function rather than  constant weighting to avoid  bias resulting from flux close to the noise level. Our analysis of galaxy orientations then uses the complex ellipticity $e=e_2+ie_2$ and the object position angle, $\theta$,  as derived from the second-order brightness moments:

\begin{equation}
e_1=\frac{Q_{xx}-Q_{yy}}{Q_{xx}+Q_{yy}},
\end{equation}
\begin{equation}
e_2=\frac{2Q_{xy}}{Q_{xx}+Q_{yy}},
\end{equation}
\begin{equation}
\theta=\frac{1}{2}\arctan (\frac{2Q_{xy}}{Q_{xx}-Q_{yy}}),
\end{equation}

where $\theta$ can be uniquely determined from the sign of $Q_{xy}$ (Fig.~\ref{fig:phi}).

\subsection{Cluster Centre Determination}\label{sec:clustercentre}

The location of the cluster centre would ideally be defined as the position of the peak of the dark-matter distribution. Since, however, maps of the dark-matter distribution are not available for the majority of our clusters, we use two alternative estimators for the cluster centre: the location  of the X-ray surface-brightness peak and the position of the BCG. We will further discuss the impact of the choice of cluster centre  in some detail in \S 4.1.2.

Seven out of the 12 clusters feature a single dominant BCG in the cluster centre; their average offset between from the X-ray surface-brightness peak is ${\sim}5^{\prime \prime}.7$ (${\sim}35$ kpc). The remaining five clusters (MACSJ0018.5+1626, MACSJ0025.4--1222, MACSJ0454.1--0300, MACSJ0647.7+7015 and MACSJ0717.5+3745) contain multiple cD-type galaxies and show clear signs of substructure. For MACSJ0018.5+1626,  MACSJ0454.1--0300 and MACSJ0647.7+7015, we choose the BCG to be the brightest galaxy within  $200$ kpc of the X-ray  peak. For MACSJ0025.4--1222 and MACSJ0717.5+3745, we choose the  brightest galaxy in the most massive substructure as identified by \citet{Bradac2008} and \citet{Ma2009}. The choice of BCGs are indicated in Figure~\ref{fig:hstimg}.

\section{Results}

\subsection{Galaxy Orientation Distribution}

PK05 and F07 find that the major axis of member galaxies in nearby clusters tend to align with their radius vector. To examine whether such a radial alignment exists in our 12 clusters at $z>0.5$, we study the distribution of the values of $\phi$, defined as the angle between the galaxy major axis and the direction toward the cluster centre (see Fig.~\ref{fig:phi}).  We stack the distributions in $\phi$ obtained for all 12 cluster fields, using a metric scale to allow us to investigate the dependence of any alignment signal with distance from the cluster core.  Note that BCGs are excluded from this analysis since their alignments reflect the surrounding large-scale structure, as discussed in \S\ref{sec:intro}.
We use spectroscopically confirmed foreground galaxies as our control sample (hereafter \textit{sample C}) since, unlike background galaxies that may be gravitationally lensed, they should not exhibit any preferred orientation.

\begin{figure}
  \includegraphics[width=0.5\textwidth]{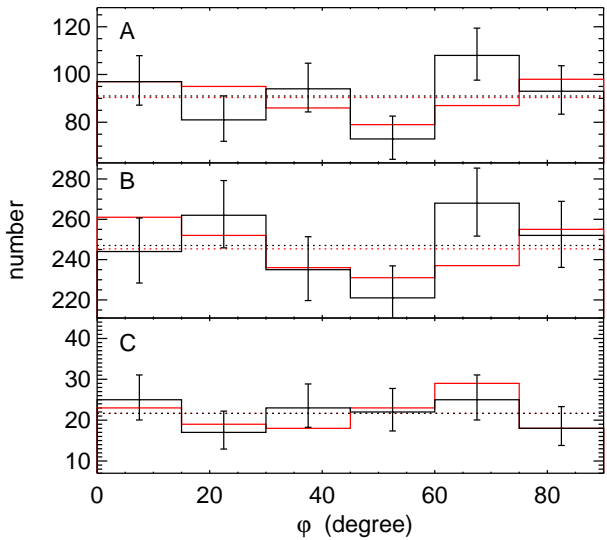} 
\caption{The distribution of galaxy orientation angles with respect to the radius vector; $\phi=0$ defines perfect radial alignment. The black and red solid lines represent the distributions obtained when adopting the X-ray brightness peak or the position of BCG as the cluster centre, respectively. Error bars assume Poisson statistics, and the dotted lines represent the average number of galaxies per  bin. In the top-left corner of each panel we denote the respective subsample (A: spectroscopically confirmed members; B: colour selected members; C: spectroscopically confirmed foreground galaxies).}
\label{fig:orientation}
\end{figure}

The distribution of galaxy orientations within the central 500 kpc (corresponding to the largest circle enclosed within the ACS images of all 12 target clusters) is shown in Figure~\ref{fig:orientation} for samples A and B. We find no tendency for radial alignment for either sample; a Kolmogorov-Smirnov test (KS test) finds the observed distributions to be consistent with a uniform parent distribution at a probability of 31 (98)\% and 62 (97)\%, respectively, when the position of the X-ray brightness peak (the BCG) is adopted as the cluster centre. As expected, the distribution of galaxy orientations for our control sample (C) is also consistent with random at a probability of 99 (99)\%. Computing the average alignment angle $\langle \phi \rangle$ (where$\langle \phi \rangle < 45^{\circ}$ indicates a preferential radial alignment), we find values of $\langle \phi \rangle$ of $45^{\circ}.45 \pm 1^{\circ}.13$ ($44^{\circ}.69 \pm 1^{\circ}.15$), $45^{\circ}.21 \pm 0^{\circ}.68$ ($44^{\circ}.55 \pm 0^{\circ}.69$), and $44^{\circ}.80 \pm 2^{\circ}.26$ ($45^{\circ}.64 \pm 2^{\circ}.29$) for sample A, B, and C.

\begin{figure}
  \includegraphics[width=0.5\textwidth]{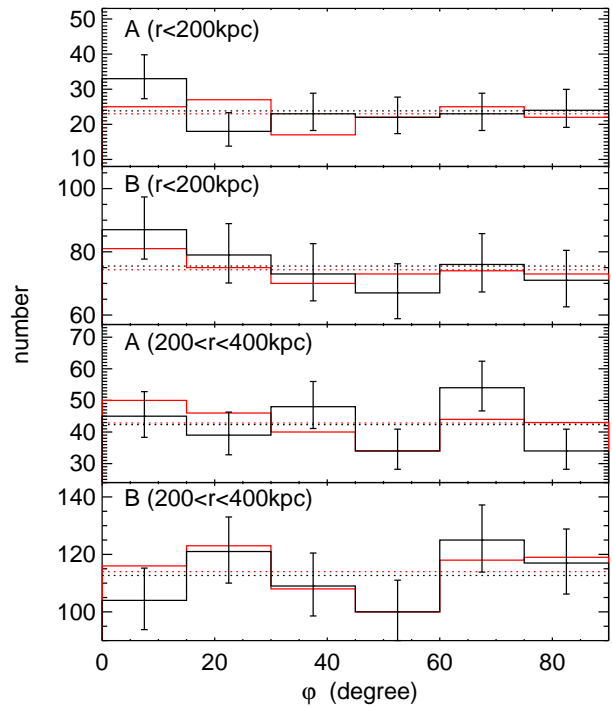} 
\caption{Same as Figure~\ref{fig:orientation} but with the galaxy sample split according to distance from the cluster centre as indicated in the top left corner of each panel.} 
\label{fig:orientationradius}
\end{figure}

\subsubsection{Dependence on Cluster Member Properties}

Although no alignment signal is detected for the largest statistical sets of cluster members compiled for our  sample, we need to bear in mind the possibility that radial alignments could be present for a subset of the cluster galaxies but are diluted to the level of insignificance when the entire galaxy sample is considered. Motived by the findings of F07, we thus examine if any  radial alignment can be found as a function of galaxy colour or distance from the cluster centre.

We first divide the samples A and B into two subsets, one including all galaxies within $200$ kpc of the cluster centre, the other comprising all galaxies within an annulus from $200$ to $400$ kpc.  A mild trend in favour of radial alignment  is  found in the central $200$ kpc region for both spectroscopic and colour-selected cluster members when the position of the X-ray surface-brightness peak is adopted as the cluster centre (Figure~\ref{fig:orientationradius}), yet a KS test still yields low probabilities of 77 and 78 \% for the observed distributions deviating from uniformity. 
Similarly, the average alignment angle is measured to be $\langle \phi \rangle =$ $43^{\circ}.43 \pm 2^{\circ}.26$ and $43^{\circ}.47 \pm 1^{\circ}.24$ for sample A and B.
No trends at all are observed when the position of the BCG is adopted as the cluster centre. No deviation from uniformity is detected in the annulus between $200$ and $400$ kpc, regardless of how cluster membership is defined. 
The slight discrepancy between the results obtained in these two radial bins may be the result of a mild dependence of the strength of radial alignments on distance from the cluster centre.

\begin{figure}
  \includegraphics[width=0.5\textwidth]{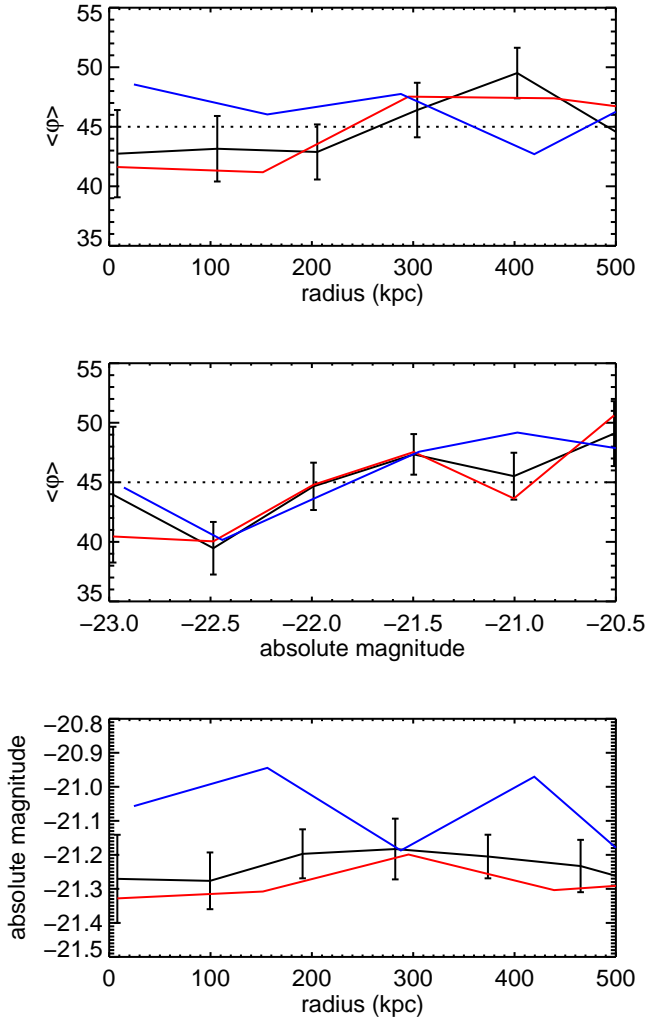} 
\caption{Upper panel: $\langle \phi \rangle$ as a function of radius. 
Middle panel:  $\langle \phi \rangle$ as a function of absolute magnitude.
Lower panel:  absolute magnitude as a function of radius.
The X-ray brightness peak is assumed as the cluster centre in this plot.
In each panel, the whole spec sample of cluster members is in black, red and blue indicate red and blue galaxy members, respectively.}
\label{fig:gpdep}
\end{figure}

Using again the average alignment angle $\langle \phi \rangle$ as a measure of  deviations from a random distribution, we test for any dependence on galaxy colour by splitting our spectroscopically confirmed sample of cluster members into two subsamples of red and blue galaxies, where ``red" is defined as lying within $2\sigma$ of the colour of the cluster red sequence (see \S\ref{sec:membership}). The results are shown in Fig.~\ref{fig:gpdep} (top panel) as a function of distance from the cluster centre. As noted before, a mild trend of increasing radial alignments with decreasing cluster-centric distances is observed, as evidenced by the fact that $\langle \phi \rangle < 45^{\circ}$ in the first three radial bins (bin size 100\,kpc);  however, only red galaxies contribute to the signal.  Although statistically insignificant, these trends are qualitatively consistent with the findings of F07.

In addition, we test for any dependence on galaxy luminosity. A similarly mild, but statistically more significant trend is found for the dependence of $\langle \phi \rangle$ on galaxy luminosity (Figure~\ref{fig:gpdep}; middle panel) in the sense that only the most luminous red galaxies show a tendency for being radially aligned. This trend is likely to be independent from the slight increase of $\langle \phi \rangle$ toward the cluster core region, since no obvious correlation is found between a galaxy's luminosity and its distance from the cluster centre.

\subsubsection{Dependence on Cluster Relaxation State}\label{sec:relax}

\begin{figure}
  \includegraphics[width=0.5\textwidth]{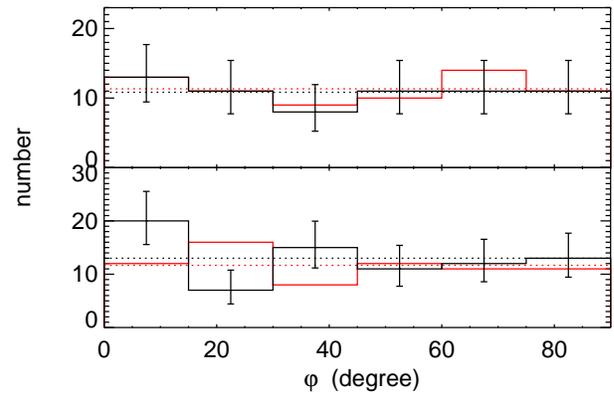} 
\caption{As Figure~\ref{fig:orientation} but showing the distribution of orientation angles for two subsets of sample A:  the six clusters classified with Morphology code $= 1 \&2$ (top panel) and the six clusters classified with Morphology code $= 3 \&4$ (bottom panel). Both plots only use data from the central $200$ kpc.}
\label{fig:orientationMorCode}
\end{figure}

As a further test of whether subsets of our sample exhibit alignments that are diluted when the full dataset (Fig.~\ref{fig:orientation}) is considered, we examine the dependence of any signal on the relaxation status of the host cluster.  Ideally, we would like to compare the alignment signal in virialized clusters with that found for morphologically disturbed, i.e., dynamically younger systems, and explore whether the strength of any alignments evolves as the host cluster approaches full relaxation after a merger.  To crudely quantify relaxation state we use the visual morphological classifications of \citet{Ebeling2007} which are based on  X-ray morphology and the goodness of the optical/X-ray alignment and result in a morphology code whose values range from $1$ for fully relaxed to $4$ for extremely disturbed clusters. We here limit the analysis to spectroscopically confirmed cluster members within the central 200~kpc, i.e. the data set for which a mild trend for radial alignments is found for the full cluster sample (but only when the X-ray peak is used to define the cluster centre; see Fig.~\ref{fig:orientationradius}, top panel).  The results for clusters with morphology codes of 1 or 2 and 3 or 4 are shown in Figure~\ref{fig:orientationMorCode}; no significant difference between the two subsets is observed.

\begin{figure}
  \includegraphics[width=0.5\textwidth]{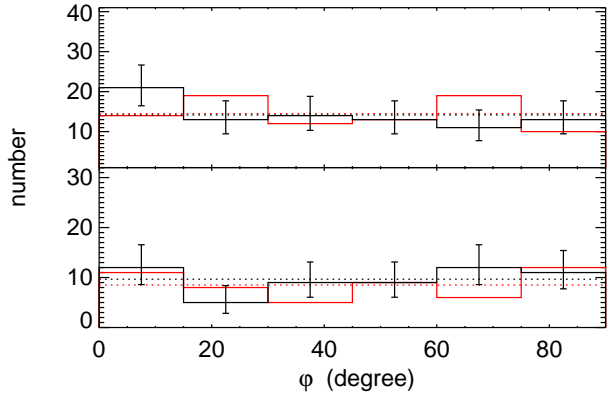} 
\caption{As Figure~\ref{fig:orientation} but showing the distribution of orientation angles for two subsets of sample A: the seven clusters featuring a single dominant BCG (top panel) and the five clusters with multiple BCG candidates (bottom panel). Both plots only use data from the central $200$ kpc.}
\label{fig:orientationMorph}
\end{figure}

Adopting an alternative, purely optical indicator of cluster relaxation state, we also examine the dependence of our findings on the presence of a single cD or multiple cD-type galaxies (see \S\ref{sec:clustercentre}). As shown in Fig.~\ref{fig:orientationMorph} (black lines), this trend originates primarily in the subsample of seven cluster featuring a single BCG, but again with very low significance (KS probability of 84\%). The presence of just one or several BCG candidates makes no difference for the distribution of orientation angle when the position of the adopted BCG is taken to define the cluster centre (Fig.~\ref{fig:orientationMorph}; red lines).

\subsection{Complex Ellipticity as a Measure of Galaxy Alignments}

Complementing our analysis based on orientation angles, we also search for galaxy alignments using the two-component complex ellipticity. Although the orientation angle and the two-component ellipticity convey similar information since they are mathematically dependent quantities, the two-component complex ellipticity has the advantage of allowing us to isolate a possible alignment signal while simultaneously testing for the presence of systematic effects in our measurement. 
Furthermore, the net alignment signal generated by a scalar gravitational cluster potential should be curl-free. Under the simplifying assumption that the cluster potential is isotropic, the curl-free and curl alignments can be decoupled by splitting the complex ellipticity $e$ into a tangential/radial component and a cross-term component, i.e.,

\begin{equation}
e_t=-\Re[|e|\exp(-2i\xi)],
\end{equation}
\begin{equation}
e_x=-\Im[|e|\exp(-2i\xi)],
\end{equation}

where $\xi$ is the angle between the galaxy radius vector and the image $x$ axis  ($\phi=|\theta-\xi|$, see Fig.~\ref{fig:phi}). 

\begin{figure}
  \includegraphics[width=0.5\textwidth]{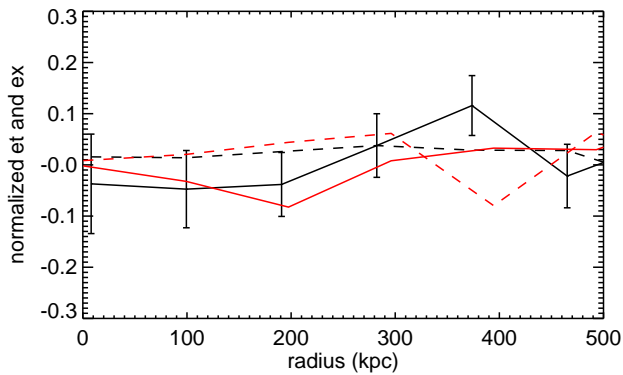} 
\caption{Variation of the two-component ellipticity with cluster-centric distance for Sample A.  The solid and dashed lines represent $e_t$ and $e_x$, respectively. Results based on the adoption of the location of the X-ray surface brightness peak (BCG) as the cluster centre are shown in black (red). To avoid cluttering, error bars are shown only for one set of results.}
\label{fig:emap}
\end{figure}

The variation of $e_t$ and $e_x$ with cluster-centric radius is shown in Fig.~\ref{fig:emap} for the sample of spectroscopically confirmed cluster members. We find a marginally significant trend toward negative $e_t$ in the central $\sim200$ kpc region, consistent with the faint radial alignment signal detected in the orientation-angle distribution within the same radial range (Fig.~\ref{fig:gpdep}, top panel). No obvious trend is observed for the cross term $e_x$, suggesting that systematic errors (other than the choice of cluster centre) are negligible.

\subsection{Possible Systematic Biases and Uncertainties}

\subsubsection{Shape Measurements}

\begin{figure}
 \centering
  \includegraphics[width=0.45\textwidth]{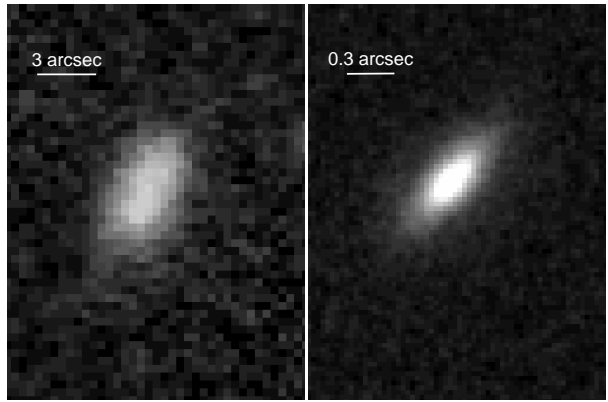} 
\caption{A comparison between the angular resolution and pixel sampling typical of SDSS ($r^{\prime}=18.5$ mag, \textit{left}) and HST (F814W$=23$ mag, \textit{right}) images of galaxies. Note the difference in angular scale covered by either image.}
\label{fig:sdss}
\end{figure}

The high quality  in terms of both angular resolution and depth  of the HST/ACS images available for our sample yields excellent pixel sampling and thus allows a much more robust shape determination than the groundbased data used in previous studies (see Fig.~\ref{fig:sdss} for an illustration). Especially the spectroscopically confirmed cluster members (which tend to be brighter) are exquisitely resolved. In this section, we examine the robustness of our galaxy shape measurements under different weighting schemes ($r_g$ in \S 3.2) and detection thresholds. 

Throughout our analysis we use weights parametrized by $r_g={\rm FWHM}/\sqrt{8\ln 2}$ as provided by SExtractor. An alternative weighting scheme, given by $r_g={\rm FWHM}$,  has been proposed by \citet{Schrabback2007}. In order to quantify the impact of the exact choice of $r_g$ on our results,  we measure the position angle of ${\sim}1000$ galaxies in one of our cluster fields (MACS\,J2129.3--0741) for either value of $r_g$. We find the average difference of $\langle\Delta\theta\rangle=-0^{\circ}.13 \pm 1^{\circ}.88$  between the derived position angles to be negligibly small compared to the bin size used by us for the orientation-angle distribution (Figs.~\ref{fig:orientation},\ref{fig:orientationradius},\ref{fig:orientationMorCode},\ref{fig:orientationMorph}) and conclude that our analysis is not sensitive to the details of the adopted weighting scheme.
 
The impact of the chosen detection threshold can, in principle, be substantial but depends greatly on the weighting function. For the Gaussian weighting function applied by us in the shape measurement process, the resulting position angle varies by only about $1^{\circ}$ when the detection threshold is raised from $5 \sigma$ to $20 \sigma$. This effect too is thus negligible for the purposes of this study.

\subsubsection{PSF Anisotropies across the Field of View}

The anisotropy of the HST Point Spread Function (PSF) is known to be a potentially significant source of systematic errors for studies involving galaxy shape measurements (e.g., weak-lensing analyses). The PSF varies across the ACS field of view and is also a function of time due to focus changes caused by ``thermal breathing" of HST in orbit \citep{Rhodes2007}. Since the cluster galaxies in our sample are well resolved we expect these PSF variations to have little effect on our study. We nonetheless test this expectation by obtaining a quantitative estimate of the variations in position angle introduced by PSF anisotropies. 

We use MACS\,J2129.3--0741 as a test field since it contains ${\sim}80$ stellar detections which allow us to sample the PSF across the field and fit position-dependent PSF models. As the ACS/WFC images of MACS\,J2129.3--0741 were taken at six  dithered positions with an integration time of $\sim 750$ seconds each, the PSF model for this field will have to take into account both spatial and temporal PSF variations. We obtain a series of  models for  focus values ranging from $-10$ to 5 $\mu$m using the IDL {\sc Tiny Tim} package \citep{Rhodes2007} based on the {\sc Tiny Tim} HST PSF modeling software \citep{Krist1995}. For each individual exposure  we find the best telescope focus by comparing the model predictions with the shape of the stellar detections in the field, and then create and stack the PSF models for these six best-fit focus values according to the dither pattern used in our observations.

Having thus obtained a PSF model for MACS\,J2129.3--0741, we apply PSF corrections using the method developed by \citet{Kaiser1995} and \citet{Hoekstra1998} which measures the response in complex ellipticity to convolution with a small anisotropic kernel. The correction term in the complex ellipticity of individual galaxies depends on their smear polarizability tensor, as defined in \citet{Kaiser1995}, and on the  second-order moments of the PSF model at each position. The resulting correction factor for the galaxy position angle is small though ($\langle\Delta\theta\rangle=1^{\circ}.57$) for the, in general, bright and well resolved galaxies in our sample. We conclude that PSF corrections have a negligible effect on the shape parameters measured for the galaxies in our cluster fields.

\section{Discussion}
Evidence of radial galaxy alignments is sparse at best for the 12 very massive clusters at $z>0.5$ investigated here. A marginal trend favouring radial alignment is found within the central $200$ kpc region around the peak of the X-ray surface brightness, but becomes entirely insignificant once the uncertainties in the determination of the cluster centre are taken into account. Within a larger region encompassing the central $500$ kpc, the $545$ spectroscopically confirmed cluster members in our sample feature an average alignment angle of $\langle \phi \rangle=45^{\circ}\!\!.38\pm1^{\circ}\!\!.14$, which is perfectly consistent with random. Our findings for clusters at $z>0.5$ are thus in conflict with the prominent radial galaxy alignments reported by PK05 and F07 for nearby clusters.  Specifically, PK05 find $\langle \phi \rangle=42^{\circ}\!\!.79\pm0^{\circ}\!\!.55$  for $2200$ spectroscopically confirmed cluster members within the central 2 Mpc of the centres of clusters at $z\sim 0.1$. We note that this discrepancy can not be attributed to statistical uncertainties: the signal reported by PK05 would still be significant at the $2 \sigma$ confidence level if their sample size were reduced by a factor of $4$ to $550$ objects. The complete lack of evidence of a net radial galaxy alignment for our $z>0.5$ sample within a much smaller distance from the cluster centres  is particularly puzzling if radial alignments are indeed strongest near the cluster cores, as suggested by both our results and F07. In this section we discuss differences between our work and previous studies that might explain the discrepant results.

We first note that the mass range probed by our cluster sample is considerably different from that used by F07. MACS, by design, selects the most X-ray luminous clusters, and the ones selected for this work feature total masses in excess of $5 \times 10^{14} M_{\odot}$ \citep{Bradac2008,Smith2009,Mantz2010,Limousin2011}; by contrast, the sample compiled by F07 using a group finder algorithm covers a much wider mass range of  $5 \times 10^{12} M_{\odot} < M_{virial} < 5 \times 10^{14} M_{\odot}$.  
The clusters in the sample of PK05, however, are X-ray selected just like ours and tend to be more massive than those studied by F07, and yet these two studies find consistent radial alignment at low redshift. We conclude that the lack of alignments observed by us at $z>0.5$ is unlikely to be caused by a difference in cluster masses.

\begin{figure}
  \includegraphics[width=0.5\textwidth]{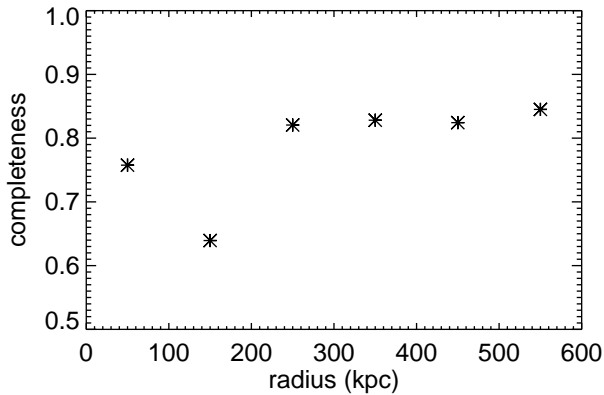} 
\caption{Spectroscopic completeness of cluster members with $m_{\rm F814W}<21.5$ as a function of clustercentric radius for our sample of 12 MACS clusters at $z>0.5$.}
\label{fig:comp}
\end{figure}

Another possible systematic difference between ours and previous work is the completeness of the spectroscopic follow-up observations of cluster galaxies. Spectroscopic coverage tends to be inhomogeneous across clusters due to conflicts between slits or fibers specifically in the dense cluster cores. Since the strength of any net radial alignment of cluster galaxies may depend on their distance from the cluster centre, spectroscopic completeness as a function of radius needs to be taken into account when comparing results obtained for different datasets. Extensive spectroscopic follow-up observations of our 12 MACS clusters result in an overall completeness of about 80\% down to F814W$=21.5$ mag. As shown in Fig.~\ref{fig:comp} the spectroscopic completeness above this limiting magnitude varies by only $\sim 15\%$ from the central 200 kpc to the edge of the ACS field of view. Correcting for relative incompleteness of the spectroscopic coverage does not change our finding that no significant radial alignment is detected within the central $500$ kpc. Although a comparison with PK05 is not possible since no information of this kind is provided for their sample, we cannot imagine a selection bias in the spectroscopic survey underlying the PK05 dataset that would explain the stark discrepancy between their results at $z\sim 0.1$ and ours at $z>0.5$.

Differences between our results and those in the literature could also be caused by biases affecting the measurement of galaxy shapes and orientations. We examine possible systematics in our analysis of HST/ACS data in \S 4.3, and find that  both instrumental and measurement systematics are negligible. Instrumental systematics should also be negligible for the datasets used by PK05 and F07 since the SDSS is conducted in drift scans, paying multiple visits to each location. Systematic errors, however, can be expected to be relevant for shape measurements based on groundbased SDSS data, owing to poor seeing and insufficient pixel sampling. 

Other potential biases inherent to alignment studies based on SDSS data are investigated by \citet{Hao2011}. Using SDSS DR7 data, they find that the radial alignment signal correlates with the apparent magnitude of BCGs but not with their absolute magnitude, causing them to conclude that such alignments are created by systematic errors in the galaxy orientation measurements caused by contamination from diffuse light of BCGs. While such a bias appears plausible in principle,  it remains unclear how diffuse light from the BCG can affect the measurement of galaxy orientations out to the large radii over which highly significant radial alignment is reported by PK05 and F07 (2 Mpc and at least $0.5 R_{\rm virial}$, respectively). Furthermore, since the control sample used by PK05 consists of field galaxies in cluster fields, at least some radial alignment trend should be present in their control sample too, if the measurement is indeed biased by diffuse light from BCGs. 
To summarize, the systematics affecting the results obtained from groundbased observations are still controversial and need further investigation.

Finally, a possible physical explanation for the differences in alignment strength between our $z>0.5$ sample and samples comprising nearby clusters is that radial alignment  evolves dramatically with redshift. If radial alignment is the result of interactions of cluster galaxies with local tidal fields, cluster members newly accreted from the field population may not have been orbiting the cluster centre long enough for tidal torquing to have a measurable effect. This effect can be expected to be stronger in younger clusters, i.e., at higher redshift. Indeed, clusters at higher redshift are known to be less relaxed than local ones \citep{Mann2011}. Although we find no correlation between cluster morphology and radial galaxy alignments  within our sample (see \S\ref{sec:relax}),  perturbations from minor or major mergers expected to be more common and frequent at larger lookback times may well prevent a radial alignment signal from reaching a significant level in distant clusters.

\section{Summary and Conclusions}
Using high-quality HST/ACS imaging data, we have performed a search for galaxy alignments within the galaxy population of 12 very X-ray luminous clusters at $z>0.5$ that constitute a statistically complete subset of the MACS cluster sample. We find no evidence of a net alignment within 500 kpc of the cluster centres ($\langle \phi \rangle=45^{\circ}\!\!.38\pm1^{\circ}\!\!.14$, based on 545 spectroscopically confirmed cluster members). Within a radius of 200 kpc of the cluster centre (defined as the location of the peak of the X-ray surface brightness), the distribution of galaxy orientations shows a mild, but statistically insignificant trend favouring radial alignments (according to a KS test, the probability of the observed distribution being different from a uniformly distributed parent sample is less than 80\%). This slight alignment trend is found to originate in red galaxies near the cluster centre. If, instead, the location of the BCG is chosen as the cluster centre, the significance of any deviation from uniformity in the alignment angles is even lower. We find no difference regarding  galaxy orientations between two subsets of our cluster sample designed to separate the more relaxed from the more disturbed systems.

In order to test for systematic errors in our analysis, we examine the distribution of both components of the complex ellipticity; as expected for curl-free alignment signals generated by a scalar gravitational cluster potential, the cross term vanishes at all clustercentric radii. We also verify the negligible impact of systematic uncertainties and biases caused by the variability of the HST point-spread function (both temporal and across the ACS field of view), and by our choice of weighting function and detection threshold when measuring galaxy ellipticities and orientations using SExtractor.

The absence of significant radial galaxy alignments in our $z>0.5$ clusters stands in stark contrast to the prominent radial alignment reported for nearby clusters. Specifically, we find no significant radial alignment of cluster galaxies within the central $0.5$ Mpc for clusters at $z>0.5$, whereas a strong alignment signal is detected out to $2$ Mpc for clusters at $z\sim 0.1$. We examine several systematic effects that could cause this discrepancy and rule out differences in the selection of clusters or cluster members. The most plausible explanations are (1) that the radial alignment evolves dramatically with cluster redshift, in the sense that the signal is weaker in dynamically younger systems, or (2) that the signal observed in nearby clusters is at least partly spurious and the result of measurement biases in the low-resolution SDSS data used by the respective studies. Both of these explanations are testable and will be examined in a future, extended study.

\section*{Acknowledgments}
HE gratefully acknowledges financial support from STScI grant GO-09722. We are indebted to David Donovan, Wendy Everett, and Maria Pereira for contributions during the early phase of this project.

\bibliographystyle{mn2e}
\bibliography{ga}

\end{document}